\documentstyle[12pt]{article}

\def\beq{\begin{equation}}
\def\eeq{\end{equation}}
\def\beqa{\begin{eqnarray}}
\def\eeqa{\end{eqnarray}}
\def\nn{\nonumber}
\newcommand{\eq}[1]{(\ref{#1})}
\def\delt{\partial_{\theta}}
\def\delx{\partial_x}     
\newcommand{\ra}{\rightarrow}
\newcommand{\NP}[1]{ {\it Nucl.{}~Phys.} {\bf #1}}
\newcommand{\PL}[1]{ {\it Phys.{}~Lett.} {\bf #1}}

\newcommand{\PRL}[1]{ {\it Phys.{}~Rev.{}~Lett.} {\bf #1}}
\newcommand{\ZPC}[1]{ {\it Z.{}~Phys.} {\bf C#1}}
\newcommand{\PTP}[1]{ {\it Prog.{}~Theor.{}~Phys.} {\bf #1}}

\newcommand{\MPL}[1]{ {\it Mod.{}~Phys.{}~Lett.} {\bf #1}}
\newcommand{\IJMP}[1]{ {\it Int.{}~Jour.{}~Mod.{}~Phys.} {\bf #1}}
\newcommand{\JP}[1]{ {\it J.{}~Phys.} {\bf #1}:\  Math.{}~Gen.{}~}
\newcommand{\LMP}[1]{ {\it Lett.{}~Math.{}~Phys.} {\bf #1}}
\newcommand{\CMP}[1]{ {\it Comm.{}~Math.{}~Phys.} {\bf #1}}
\newcommand{\JMP}[1]{ {\it J.{}~Math.{}~Phys.} {\bf #1}}
\begin{document}

\topmargin 0pt
\oddsidemargin 5mm
\evensidemargin 5mm

\begin{titlepage}
\setcounter{page}{0}
%
\begin{flushright}
NBI-HE-96-66\\
OWUAM-017 \\
INS-Rep-1173\\
hep-th/9706176\\
\end{flushright}
\vspace{13mm}
\begin{center}
{\Large   Notes on Curtright-Zachos Deformations of $osp(1,2)$ \\
and Super Virasoro Algebras } 
\vspace{15mm}

{\large Naruhiko Aizawa$^{a)}$
{}~\footnote{{}~aizawa@appmath.osaka-wu.ac.jp  }}   
{\large Tatsuo Kobayashi$^{b)}$
{}~\footnote{{}~kobayast@ins.u-tokyo.ac.jp  }}   
%
%
{\large Haru-Tada Sato$^{c)}$
{}~\footnote{{}~Fellow of the Danish Research Academy,\,\,\, 
 sato@nbi.dk}}\\

\vspace{5mm}
$^{a)}${\em Department of Applied Mathematics, Osaka Women's University\\
             Sakai, Osaka 590, Japan}\\
$^{b)}${\em Institute for Nuclear Study, University of Tokyo\\
            Midori-cho, Tanashi, Tokyo 188, Japan }\\
$^{c)}${\em The Niels Bohr Institute, University of Copenhagen\\
            Blegdamsvej 17, DK-2100 Copenhagen, Denmark}\\

\end{center}
\vspace{5mm}

\begin{abstract}
Based on the quantum superspace construction of $q$-deformed algebra, 
we discuss a supersymmetric extension of the deformed Virasoro algebra, 
which is a subset of the $q$-$W_{\infty}$ algebra recently appeared 
in the context of two-dimensional string theory. We analyze two types 
of deformed super-Virasoro algebra as well as their $osp(1,2)$ 
subalgebras. Applying our quantum superspace structure to conformal 
field theory, we find the same type of deformation of affine $sl(2)$ 
algebra.                                                               
\end{abstract}

\vspace{1cm}

\end{titlepage}
\newpage
\renewcommand{\thefootnote}{\arabic{footnote}}
\section{Introduction}
\indent

The idea of $q$-deformation has become useful to bring rich contents 
into physical models and has been identified as a suitable method to 
formulate a deformation of ideal physical situation. Sometimes, the 
$q$-deformation is discussed in a connection to the so-called quantum 
deformation studied in the area of mathematics and integrable 
systems. The latter strictly requires non-trivial structures of Hopf 
algebra and Yang-Baxter equation \cite{QG}. However, in some 
situation which does not need the non-trivial mathematical structures, 
it is much convenient to use a simple $q$-deformation. The 
typical example of such objects is a $q$-oscillator, which has been 
used systematically everywhere as a natural consequence that the
(undeformed) harmonic oscillator is a fundamental element in a wide 
range from quantum mechanics to field theory. 
The $q$-oscillators are used not only as a convenient realization of 
quantum deformation \cite{qos}, but also for the purpose of getting 
closer to experimental consistency and more general situation 
\cite{Pauli}-\cite{Fuji}. 
Every deformation contains its original model by construction as 
the special case $q=1$.

{}~For a curious example of applications of $q$-oscillator, we would 
like to mention a realization of the $q$-deformed Virasoro algebra 
which is proposed by Curtright and Zachos (CZ) \cite{CZ}. Recently, 
Jevicki et. al. found a realization of one of $q$-deformed $W_{\infty}$ 
algebras \cite{zha} in $c=1$ matrix model using a finite 
representation of ($q$-deformed) fermion oscillators \cite{jev}. 
It is well known that the $c=1$ matrix model is expressed by 
two-dimensional free fermion theory in the double scaling limit 
taken by letting the size of matrix(number of fermions) go to infinity 
and the lattice spacing to zero. Hence, as far as keeping the number of 
fermions finite, the model corresponds to a discretized system, and we  
encounter the $q$-$W_{\infty}$ algebra in the discretized theory 
with the connection to the fact that the free fermion possesses the 
$W_{\infty}$ symmetry, which is related to the Virasoro constraints 
on physical states. This $q$-deformed $W_{\infty}$ algebra includes 
the $q$-deformed Virasoro algebra. 

In this paper, we consider a supersymmetric extension, of the 
original work of Curtright and Zachos, introducing a quantum 
superspace approach \cite{QSST}. Generally speaking, commutation 
relations among coordinates and derivatives in quantum space are 
systematically governed by a quantum group 
$R$-matrix \cite{WZ}. We do not know yet what deformed (CZ type) 
super Virasoro algebra could be realized either in a super matrix 
model (physical) approach \cite{fermi} or even in a usual 
$q$-deformation (mathematical) approach. 

To study the deformation here, we prefer to take a quantum superspace 
approach rather than super $q$-oscillator's one from the following 
background thoughts. {}~First of all, the non-commutativity of quantum 
(super)space is a likely option of unusual spacetime structure 
which would be anticipated in an extremely high energy situation 
--- it might be a non-archimedian geometry \cite{arch} or the 
foam-like structure of spacetime at the Planck scale. The 
$q$-deformation ($q$-regularization \cite{qregular}, in other words) 
of field theory could provide a natural UV cut-off, and we thus 
suppose to have a model of non-divergent theory, in particular, of 
quantum gravity. There exist some models along this line 
\cite{qgauge},\cite{qscalar}. 

Secondly, the $q$-oscillator is nothing but a literal deformation 
of the harmonic oscillator, and we have not got any answer to the 
question what fundamental or guiding principle of deformation is 
behind the $q$-oscillator. A possible answer would seem to lie in 
the world of quantum space. Quantum space geometry is determined 
from covariance under the transformation by a quantum matrix group, 
whose matrix entries are non-commutative numbers. Some $q$-oscillator 
shows the similar behaviour as the quantum group covariant basis 
\cite{cova}. This suggests that the $q$-oscillator is realized on the 
quantum space. In this sense, the quantum space approach is rather 
fundamental than the $q$-oscillator approach is, if we assume 
the deformation structure of spacetime.
 
Our present aim is to present a method how to obtain a supersymmetric 
extension of CZ algebra at the level of operator algebra of quantum 
superspace. Although this note does not make a complete abstract 
study of the deformed superalgebra, we intend to discuss the algebras 
satisfied by the {\it real} operators defined on quantum superspace.

The contents of this paper are as follows. 
In sect.2, we explain the notations and formulas of our quantum 
superspace \cite{KU}. We present also a realization of the 
$q$-deformed Virasoro algebra (we call CZ algebra) on the quantum 
superspace. {}~For the purpose of finding universal structures, we 
have to examine various choices and possibilities. In this sense, we 
study two types of supersymmetric extension of CZ algebra in sect.3. 
In each case, the generators of superalgebra will be given by the set 
of $B_n$, $F_n$ and $G_r$. In sect.4, we show that $B_n$ corresponds 
to the CZ generator $L_n$, while $F_n$ corresponds to a non-linear 
deformation term made of bilinear combination of $G_r$. According to 
this structure, we summarize the general prescription how to 
construct a supersymmetric extension of CZ algebra. Sect. 5 contains 
for reference a short observation of a map relation, through 
$q$-oscillator representations, between one of our deformed $osp(1,2)$ 
subalgebras and the Drinfeld-Jimbo type deformation \cite{osp}. 
In sect.6, we append an example of application of quantum superspace 
to a physical system. Applying an algebraic structure of the quantum 
superspace to free field representation of conformal field theory, we 
derive the CZ type deformations of $sl(2)$ and affine $sl(2)$ current 
algebras. Sect.7 is conclusion.

\section{Quantum superspace and CZ algebra}
\setcounter{equation}{0}
\indent                       


We deform the centerless $N=1$ super Virasoro algebra 
(for example \cite{FMS})
\beqa
&& [{\cal L}_n,{\cal L}_m]=(n-m){\cal L}_{n+m}, \label{nondefll}  \\
&& [{\cal L}_m,{\cal G}_r]=
                 ({1\over2}m-r){\cal G}_{m+r},   \label{nondeflg} \\
&& \{{\cal G}_r,{\cal G}_s \}=2{\cal L}_{r+s},   \label{nondefgg} 
\eeqa
and its subalgebra $osp(1,2)$ using coordinates and derivatives on 
quantum superspace. Our method is to deform a differential operator 
realization through replacing the undeformed differential operator 
with quantum superspace's one. The same method was applied to 
eq.\eq{nondefll} using a bosonic quantum space to obtain the CZ 
algebra \cite{hass1}. In this paper, we consider the following two 
types of undeformed realization: we call the {\it first type} for 
\beq
{\cal L}_n=-x^n( {n+1\over2}\theta\delt + x\delx ), \quad 
{\cal G}_r= x^{r+1/2}(\delt -\theta\delx ),         \label{first}
\eeq
and the {\it second type} for 
\beq
{\cal L}'_n = -x^{n+1} \delx -{n \over 2} x^n \theta \delt, \quad
{\cal G}'_r =  x^r(\delt -x \theta \delx ), \quad   \label{second}
\eeq
where these $x$ and $\theta$ are obviously undeformed. 

In what follows, we suppose that each of $x$ and $\theta$ etc. 
expresses quantum superspace notation. The commutation relations of 
our quantum superspace are \cite{KU} defined by 
\beqa
&&(\theta)^2=(\partial_\theta)^2=0, \quad x \theta =  q\theta x,
\quad \delx \delt =  q^{-1}\delt \delx,        \nn   \\
&&\delx x=1+q^{-2} x \delx, \quad \delt \theta 
           = 1- \theta \delt +(q^{-2}-1)x \delx, \label{qspace} \\
&&\delx\theta=q^{-1}\theta\delx, \quad \delt x=q^{-1}x\delt. \nn
\eeqa
Since we need some amount of calculation as expected from the above 
awkward commutation relations, we first list up useful formulas 
for convenience ($n\in {\bf Z}$ or ${\bf Z}+1/2$) 
\beqa
& & \delx x^n = q^{-2n} x^n \delx + q^{-n+1} [n] x^{n-1}, \\\
& & \theta \delt x^n = q^{-2n} x^n \theta \delt, \\
& & \theta \delt \delx = q^{2} \delx \theta \delt, 
\eeqa                                               
and introduce the scaling operator $\mu$
\beq
\mu \equiv \partial_x x-x\partial_x=1+(q^{-2}-1)x\partial_x,
\eeq
which satisfies
\beqa
&&\mu x=q^{-2}x \mu, \qquad \mu \partial_x =q^2 \partial_x \mu, \\
&&[\mu,\theta]=[\mu,\delt]=0.
\eeqa
The following are also useful
\beqa
\delt\theta & = & \mu-\theta\delt, \\
\delx x^n & = & - q^{-2n+1}x^{n-1}{\mu-q^{2n}\over q-q^{-1}}, 
\qquad n\not=0,\\
(\theta\delt)^n & = & \mu^{n-1}\theta\delt. 
\eeqa                                                                  

We introduce the following fundamental deformed operators, which will 
form the bases of super CZ algebra:
\beqa
B_n=-q^{-1}x^{n+1}\delx,\\ 
F_n=-g_nx^n\theta\delt,  
\eeqa
where $g_n$ is a $q$-dependent constant (which will be determined 
later) and supposed to be reduced to the coefficient of $\theta\delt$ 
in \eq{first} or \eq{second} as $q\ra1$. These satisfy
\beqa
&&[ B_n , B_m ]_{(m-n,n-m)} = [n-m]B_{n+m}, \label{bb}  \\
&&[ F_n , F_m ]_{(m-n,n-m)} = 0,            \label{ff}  \\
&&[B_n ,F_m ]_{(m-n,n-m)} = -q^{-n}[m]F_{n+m}. \label{bf} \\
&&\mu^{-\alpha} B_n \mu^{\alpha} = q^{2n\alpha} B_n,\qquad
  \mu^{-\alpha} F_n \mu^{\alpha} = q^{2n\alpha} F_n,  
\eeqa
where $[n]=(q^n-q^{-n})/(q-q^{-1})$ and 
\beq 
            [A,B\}_{(a,b)}=q^{a} AB \pm q^{b} BA.
\eeq                                 
Eqs.\eq{bb} and \eq{ff} are the CZ algebra \cite{CZ} and the 
deformed $U(1)$ Kac-Moody algebra, which are derived from 
$gl(\infty,{\bf C})$ in \cite{hass2} (To be more precise, we 
have to change $q\ra q^{-1}$ and $B_n\ra q^{-n}B_n$).
Of course, $F_n$ can be understood as a deformation of $U(1)$-element 
which plays a part of $N=2$ super Virasoro algebra. 
If we require that the combination as well 
\beq
               L_n = B_n + F_n            \label{ln}
\eeq
satisfies the CZ algebra \eq{bb}, i.e.,
\beq
q^{m-n} L_n L_m - q^{n-m} L_m L_n = [n-m] L_{n+m}, \label{czalgebra}
\eeq                                                                   
$g_n$ should be a solution of 
\beq
      q^{-n} [m] g_m - q^{-m} [n] g_n = [m-n] g_{n+m}. \label{keisuu}
\eeq
The general solution of \eq{keisuu} is given by 
\beq
g_n = aq^{-2n} + b,   \label{gnsoln}
\eeq
where $a$ and $b$ are $n$-independent arbitrary constants. 
{}~For the associativity of algebra, $L_n$ 
should satisfy the following relation \cite{AS}
\beq
[m-l][L_n,L_{m+l}]_{(m+l,-m-l)} + \mbox{cyclic perms.} =0.
\eeq
This can be checked by using our $L_n$ \eq{ln}. This is nothing but 
a consequence of associativity of the quantum superspace 
differential operators. The whole space spanned by $L_n$ should be 
divided by this relation. 

\section{Deformed $osp(1,2)$ and super CZ algebras}
\setcounter{equation}{0}
\indent 

In this section, we consider the two types of deformation of $N=1$ 
super Virasoro algebra. As shall be shown, $F_n$ appears as redundancy 
in these deformed superalgebras (however it disappears when $q\ra1$). 
The appearance of $F_n$ suggests that $N=2$ 
deformation might be natural in the cases of $q\not=1$; in other words, 
our $q$-deformations of the $N=1$ algebra induce a deformed $N=2$ super 
Virasoro algebra. In this sense, the $N=1$ deformations are very much 
non-trivial. Since the decomposition of our deformed algebra into $N=2$ 
algebra is an easy exercise, we shall focus our attention on the 
$N=1$ deformed algebras. A nontrivial thing is, as shall be shown 
in sect.4, how to eliminate the $N=2$ algebra element $F_n$ from 
our deformations. 

\subsection{First type deformation}
\indent

Let us deform the first type \eq{first}. 
For the CZ generators $L_n$ defined in \eq{ln}, we choose 
\beq 
         g_n  = {1\over[2]}q^{-n}[n+1].      \label{gnsolution}
\eeq            
As to a deformation of ${\cal G}_r$, we define
\beq
G_{m-1/2} = \mu^{-1/2}x^m(\delt- \theta\delx), 
\eeq
and introduce the following compact notations for convenience's sake,
\beq
G_{m-1/2}^{(\beta)}=\mu^{-1/2}x^m(\delt-q^{2\beta}\theta\delx), 
\quad G_{m-1/2}^{(0)}\equiv G_{m-1/2}.
\eeq       
Note that we can always reduce the quantity $G_m^{(\beta)}$ 
into $G_m$ using 
\beq
G_{m-1/2}^{(\beta)}=
          q^{2m\beta}\mu^{\beta}G_{m-1/2}\mu^{-\beta}. \label{mugmu}
\eeq                                                             
Under this definition of $G_n$, the product of two $G_n$'s becomes 
\beq
G_{m-1/2}G_{n-1/2}=
-q^{n+m}x^{n+m}\delx-q^{m+1}[n]x^{n+m-1}\theta\delt.
\label{ggproduct}
\eeq
Now, the commutation relations among $B_n$, $F_n$ and $G_r$ are 
\beqa
{}[B_n,G_{m-1/2}]_{(m-n,n-m)} & = & {-q^{n-m}\mu\over q-q^{-1}}
\left( G^{(m-n/2-1)}_{n+m-{1\over2}} 
         - q^{2m-n}G_{n+m-{1\over2}}\right) \nn \\
                 & - & q^{-n}[m-n]G_{n+m-{1\over2}}, \label{bg}  \\
{}[F_n,G_{m-1/2}]_{(m-n,n-m)} & = & {1\over[2]}
{q^{n+m}\mu\over q-q^{-1}}\left(q^{n+1}G^{(-2n-1)}_{n+m-{1\over2}} 
              - q^{-n-1}G^{(1)}_{n+m-{1\over2}}\right),\label{fg}
\eeqa
and
\beq
\{G_{m-1/2},G_{n-1/2}\}_{(0,-2n)} =(1+q^{-2n})q^{n+m+1}B_{n+m-1}
+[2]q^{n+m}F_{n+m-1}.        \label{deformedgg}
\eeq
Taking the combination given by \eq{ln} with \eq{gnsolution}, we can 
obtain the alternative expression of \eq{bg} and \eq{fg}
\beqa 
[L_n,G_{m-1/2}]_{(m-n,n-m)} & =& q^{n-m-2n(n+m+1)}{[1-n]\over[2]}
               \mu^{-n} G_{n+m-1/2}\mu^{n+1}                \nn\\
                       & + & q^{-n}[n-m]G_{n+m-1/2}. \label{lmgn}
\eeqa                    
It is clear that \eq{czalgebra}, \eq{deformedgg} and \eq{lmgn} 
reproduce the super Virasoro algebra \eq{nondefll}-\eq{nondefgg} in 
the limit $q\ra1$. 

Here we remark on the closure of the algebra. As will be shown in 
sect.4, the $B_n$ and $F_n$ terms on RHS of \eq{deformedgg} can be 
understood as $L_n$ and $G_n$ terms. 
In addition, we should notice the following relation  
\beq
         \mu = 1-q^{-1}(q^{-2}-1)G_{1/2}G_{-1/2}.
\eeq
Every algebraic relation is hence expressed only by $B_n$, $F_n$ and 
$G_r$, and the super algebra is closed concerning $B_n$,$F_n$ and 
$G_r$ (and thus $L_n$ and $G_r$). Terms including $\mu$ produce 
non-linear deformation terms. 
It may be interesting to note also that 
\beqa
\delx &=& -qL_{-1}, \\
\theta\delt &=& q^{-1}[G_{1/2},G_{-1/2}].
\eeqa                           

Let us look at the $osp(1,2)$ subalgebra deformation, restricting 
$s$, $r=\pm1/2$, $n=0,\pm1$. The above situation of $B_n$ and $F_n$ 
becomes much simpler in this case. In fact, eq.\eq{deformedgg} becomes
\beq
\{G_r,G_s\}_{(s-r,r-s)}=q^{2+r+s}(q^{s-r}+q^{r-s})L_{r+s},
\eeq
which contains the independent relation as well as the 
subordination relations of $L_{\pm1}$ to $G_{\pm1/2}$:
\beqa
&& \{G_{1/2},G_{-1/2}\}_{(-1,1)} =q^2[2]L_0, \label{ospresulta} \\
&& (G_{-1/2})^2 = q L_{-1},
         \qquad (G_{1/2})^2 = q^3L_1.           \label{lgivenbyg} 
\eeqa
Eq.\eq{lgivenbyg} means that $L_{\pm1}$ are not independent elements of 
the deformed $osp(1,2)$ algebra any more. This situation is exactly 
the same as that of the standard deformation of $osp(1,2)$ \cite{osp12}. 
The remaining commutation relations following from \eq{lmgn} are the 
independent relations between $L_0$ and $G_{\pm1/2}$,
\beqa
&& [L_0,\;G_{-1/2}] ={1\over[2]} G_{-1/2} \mu, \label{ospresultb}\\
&& [L_0,\;G_{1/2}]_{(0,-2)}=
    -q^{-1}G_{1/2}+q^{-2}{1\over[2]}G_{1/2}\mu, \label{ospresultc}
\eeqa
and the subsidiary relations
\beqa
&& [L_{\pm1}, \; G_{\pm1/2}] = 0, \\
&& [L_{\pm1}, \; G_{\mp1/2}]_{(0,\pm2)} = \pm G_{\pm1/2}.
\eeqa
The deformed $osp(1,2)$ algebra which consists of three 
(anti-)commutation relations \eq{ospresulta}, \eq{ospresultb} and 
\eq{ospresultc} is generated by $L_0$ and $G_{\pm1/2}$, and 
the subordination relations \eq{lgivenbyg} exclude $L_{\pm1}$ from 
these commutation relations.

\subsection{Second deformation}
\indent                     

We repeat the similar observation on the deformation of the second type 
\eq{second}. Let us define the deformed operators
\beq
G'_r = \mu^{-1/2} x^r(\delt -x \theta \delx ), \qquad
L'_n  = B_n + F'_n,
\eeq
where $F'_n$ are defined with the following choice of $g_n$ 
\beq
        g'_n = q^{-n/2}[{n\over2}].
\eeq                            
Note $L'_n$ do not satisfy the CZ algebra \eq{czalgebra}. Instead, 
they satisfy
\beq
[L'_n,L'_m]_{(m-n,n-m)}=[n-m]L'_{n+m} -
{[n]-[m]-[n-m] \over [{n+m\over2}](q-q^{-1})}q^{-(n+m)/2}F'_{n+m},
\label{NLCZ}
\eeq
where the second term will be shown to be composed of $G_r$ as 
remarked before. The counterparts of \eq{mugmu} and \eq{ggproduct} 
are{}~\footnote{In the following sections, we express $L'_n$ 
($G'_r$) simply by $L_n$ ($G_r$) as far as no attention.} 
\beq
\mu^{-\alpha}G_r\mu^{\alpha} = q^{2r\alpha}G_r, \label{newmgm}
\eeq
\beq
G_rG_s = -q^{r+s+1}x^{r+s+1}\delx 
         - q^{r+2}[s]x^{r+s}\theta\delt,      \label{ggprod2}
\eeq
where $\mu$ is connected to $L_0$ by the relation 
\beqa
     \mu &=& 1+(q-q^{-1})L_0.
\eeqa
Eq.\eq{newmgm} seems favorable rather than the former case \eq{mugmu} 
because of no appearance of non-linear term in the commutation 
relation between $L_n$ and $G_r$ (c.f. \eq{lmgn})
\beq
[L_n,G_r]_{(r-{n\over2},{n\over2}-r)} =q^{-n}[{n\over2}-r]G_{n+r}.
\label{lgprod}
\eeq
The other relations are 
\beq
\{G_r,G_s\}_{({s-r\over2},{r-s\over2})} =q^{r+s+2}
 (q^{s-r\over2}+q^{r-s\over2}) L_{r+s},  \qquad (r=\pm s)  
\label{ggspecial}
\eeq
\beq
\{G_r,G_s\}_{(0,-2s)}=
 (1+q^{-2s})q^{r+s+2}B_{r+s}+q^2(1+q^{r+s})F_{r+s}.
 \qquad\mbox{otherwise} \label{gggeneral}
\eeq
It is clear that \eq{lgprod}, \eq{ggspecial} and \eq{gggeneral} 
are reduced to \eq{nondeflg} and \eq{nondefgg} in $q\ra1$. 
Obviously \eq{NLCZ} recovers the Virasoro algebra \eq{nondefll} when 
$q\ra1$. Thus \eq{lgprod}-\eq{gggeneral} and \eq{NLCZ} form a 
supersymmetric extension of CZ algebra in the same sense as the 
first type deformation.

The subalgebra structure of this deformed superalgebra is as follows. 
The $su(1,1)$ subalgebra deformation reads
\beqa
&& [L_n,L_0]_{(-n,n)}=[n]L_n \\
&& [L_1,L_{-1}]_{(-2,2)}=[2]L_0
-q^{-2}(q^{1\over2}-q^{-{1\over2}})[{1\over2}][G_{1/2},G_{-1/2}],
\eeqa
where $n$ can be extended to an arbitrary integer. Up to the 
non-linear terms on RHS of the above equations, these coincide 
with the $su(1,1)$ part of the CZ algebra. Again, $L_{\pm1}$ are 
given by $G_{\pm1/2}$ through the same relations as \eq{lgivenbyg}, 
and the above $su(1,1)$ parts are covered by the 
following independent (anti-)commutation relations 
\beqa
{}& &\{G_{1/2},G_{-1/2}\}_{(-{1\over2},{1\over2})}
      =q^2(q^{1/2}+q^{-1/2})L_0,  \\
{}& &[L_0,G_r]_{(r,-r)}=[-r]G_r.
\eeqa
Differently from the first type deformation, the deformed $osp(1,2)$ 
commutation relations are now linear forms. Instead, the CZ algebra 
acquires non-linear deformation terms seen in \eq{NLCZ}.
In the next section, we give the interpretation of $B_n$ and $F_n$, 
which are given by $G_r$ bilinear forms. 

\section{General prescription of $N=1$ super CZ algebra}
\setcounter{equation}{0}
\indent                     

In this section, we give a general prescription of obtaining a super 
CZ algebra. First, let us explain our idea how the bases of super CZ 
generators are chosen in the second type case (one 
can repeat exactly the same consideration for the first type 
deformation). It is notable that $L_{\pm1}$ are no more independent 
elements in the deformed $osp(1,2)$ algebra through the relations 
\eq{lgivenbyg}, but their degrees of freedom correspond to bilinear 
terms of $G_{\pm1/2}$. We formally generalize this idea to $L_n$, 
$n\not=0$, and thus we regard that our supersymmetric extension of 
CZ algebra consists of only $L_0$ and $G_r$. 
Namely, from \eq{ggspecial}, 
\beq
  L_n = q^{-n-2} (G_{n/2})^2,  \qquad n\not=0,   \label{reduce}
\eeq
we formally drop every $L_n$ ($n\not=0$) away from the commutation 
relations \eq{NLCZ}, \eq{lgprod}-\eq{gggeneral}. 
Strictly speaking, we can merely drop either odd or even modes. 
However resulting expressions obtained from \eq{reduce} with the use 
of anti-commutators of $G_r$ can be generalized to even/odd modes. 
This occurs in the undeformed case as well. As a result, 
the remaining commutation relations determine the superalgebra: 
i.e., \eq{gggeneral} and 
\beqa
&& \{G_r,G_{-r}\}_{(-r,r)}=(q^{-r}+q^{r})q^2L_0, \label{GGL}\\
&& [L_0,G_r]_{(r,-r)}=-[r]G_r.              \label{LG}
\eeqa

Now, the only question is how to interpret the $B_n$ and $F_n$ 
appeared in \eq{gggeneral} (or \eq{deformedgg} in the first type).
Fortunately, an arbitrary linear combination of $B_n$ and $F_n$ can 
be expressed in terms of the new bases
\begin{eqnarray}
&& {\cal B}_n \equiv (G_{n/2})^2 = -q^{n+1}x^{n+1}\delx
-q^{2+n/2}[{n \over 2}]x^n\theta \delt,\\
&& {\cal F}_n \equiv G_{(n+1)/2}G_{(n-1)/2} = -q^{n+1}x^{n+1}\delx
-q^{(n+5)/2}[{n -1\over 2}]x^n\theta\delt,
\end{eqnarray}
where the second equality in each equation is clear from \eq{ggprod2}. 
Obviously, these are nothing but certain recombination of $B_n$ and 
$F_n$. Therefore any closed algebra on $B_n$ and $F_n$ is a closed 
algebra on $L_n$ and $G_r$, and it acquires non-linear deformation 
terms. 

Here we comment on the general features of the super CZ algebras: 
(i) We could have used other 
elements to define ${\cal F}_n$, for instance, ${\cal F}_n = G_{n-r}G_r$. 
There are many choices of ${\cal F}_n$ and accordingly a variety of 
super CZ algebras. It is still an open question 
which choice of ${\cal F}_n$ is convenient or essential for 
$\{G_r,G_s\}$ deformation. (ii) Nevertheless, it is clear that the 
${\cal B}_n$ term gives the $L_n$ term (because of \eq{reduce}), and 
the ${\cal F}_n$ term gives a non-linear deformation term which 
disappears in $q\ra1$. (iii) The general program to construct a super 
CZ algebra is as follows. First, define $G_r$ operators using quantum 
superspace operators and calculate the $G_rG_s$ product. 
Secondly, define ${\cal B}_n$ and ${\cal F}_n$ as $G_r$ bilinear forms. 
Then we can solve $B_n$ and $F_n$ (with an appropriate $g_n$) in terms 
of ${\cal B}$ and ${\cal F}$. This trivially 
defines the CZ operators $L_n$ of \eq{ln} in terms of $G_r$.
Finally, in terms of $L_n$, $G_r$ and ${\cal F}_n$, rewrite $B_n$ and 
$F_n$ terms which still remain in the (anti-)commutation relations.   
This gives a super CZ algebra which is closed on $L_n$ and $G_r$. 
Note that even if one calculates these commutation relations 
starting with the $L'_n$ operator which is different from the 
CZ operator $L_n$, $L'_n$ can always be re-expressed in terms of 
$L_n$ and $F_n$. 

\section{$q$-oscillator representation of deformed $osp(1,2)$}
\setcounter{equation}{0}
\indent

We put a short note on the relation of our $q$-$osp(1,2)$ to the 
standard one \cite{osp12},\cite{osp}, which possesses a Hopf algebra 
structure, via relationship between the following two different 
deformed harmonic oscillators. For convenience, we use the word 
{\it q-oscillator} to stand for the relation 
\beq
\alpha\alpha^{\dagger}-q\alpha^{\dagger}\alpha=1,   \label{four}
\eeq
and {\it p-oscillator} to stand for \cite{qos} 
\beq
 a a^{\dagger} - p^{-1} a^{\dagger} a = p^N, \quad 
 [N, \; a] = -a, \quad 
 [N, \; a^{\dagger}] = a^{\dagger},      \label{one}
\eeq
with the relations
\beq
a^{\dagger}a =[N]_p,\qquad a a^{\dagger} = [N+1]_p,  \label{two}
\eeq
where
\beq
  [X]_p \equiv {p^X - p^{-X} \over p-p^{-1} },
\eeq
and we choose the value of $p$-oscillator's Casimir to be zero 
for simplicity. The standard $q$-$osp(1,2)$ algebra 
(for example see \cite{osp}) is 
\beq
  [H, \; V_{\pm}] = \pm {1\over2} V_{\pm},\qquad
  \{V_+, \; V_-\} = - {1 \over 4} [2H]_p.        \label{twopf}
\eeq
This algebra is given by the trivial renormalization of the 
$p$-oscillator \cite{map}
\beq
  V_+ = {1 \over 2 \sqrt{p^{1/2}+p^{-1/2}}} a^{\dagger}, \quad 
  V_- = {-1 \over 2 \sqrt{p^{1/2}+p^{-1/2}}} a,   \quad
  H = {1\over2} (N+ {1\over2} ).                     \label{three}
\eeq
On the other hand, our deformed $ osp(1,2) $ of the second type given 
in eqs.\eq{GGL} and \eq{LG} can be realized by the $q$-oscillator
\beq
L_0 = c_0(q^{-1}\alpha\alpha^{\dagger}+\alpha^{\dagger}\alpha),\quad 
G_{-1/2} = c \alpha^{\dagger}, \quad
G_{1/2} = c \alpha,                                  \label{six}
\eeq
where 
\beq
    c_0 = {1 \over q^{-1} + 1} q^{1/2} [{1\over2}]_q, \qquad 
    c = \sqrt{q^{5/2} [{1\over2}]_q}.
\eeq
Using the connection between these oscillators \cite{cova}
\beq
\alpha=q^{N/4}a,\qquad\alpha^{\dagger}=a^{\dagger}q^{N/4},\qquad
  p = q^{-1/2},                                     \label{seven}
\eeq
we can realize the following mapping relations on the $p-$oscillator 
Fock space
\beqa
 & & L_0 = {\displaystyle c_0 \over q^{1/2}-q^{-1/2}} 
           (2q^{2H-1} - q^{-3/2} - q^{-1/2}),          \\
 & & G_{-1/2} = 2c \sqrt{(q^{1/4}+q^{-1/4})} V_+ q^{(4H-1)/8}, \\ 
 & & G_{1/2} = 2c \sqrt{(q^{1/4}+q^{-1/4})} q^{(4H-1)/8} V_-. 
\eeqa
Thus, our $q$-$osp(1,2)$ algebra is a linearized version of 
(\ref{twopf}) in this representation. The reverse mapping is also 
found in the similar way, 
\beqa
& & H={-1\over4}\log_p{1\over2}(p^2+1-p(p^2-p^{-2}) [2]_p L_0) 
     + {1\over4},   \\
& & V_+ ={1\over2}\sqrt{p^5 [2]_p \over p^{1/2}+p^{-1/2}} G_{-1/2} 
     \left\{ {1\over2} (p^2+1-p(p^2-p^{-2})[2]_pL_0)\right\}^{-1/4},
     \\ 
& & V_- = {1\over2} \sqrt{p^5[2]_p\over p^{1/2}+p^{-1/2}} 
     \left\{ {1\over2} (p^2+1-p(p^2-p^{-2})[2]_pL_0\right\}^{-1/4} 
     G_{1/2}.
\eeqa
                                                                     
\section{Deformed affine current algebra}
\setcounter{equation}{0}
\indent                                  

This section concerns a simple analogy of conformal field theory, 
defining free fields associated with our quantum superspace. 
We have obtained the differential operator realizations of deformed 
$osp(1,2)$ algebra on the quantum superspace. We can apply this 
realization to the relation between free field and differential 
operator representations \cite{BO}; i.e.
\beqa
&& J^+ = -\beta,    \\
&& J^- = \beta\gamma^2+\alpha_+\gamma\partial\phi+\gamma cb 
          -k\partial\gamma + (k+1)c\partial c,   \\
&& J^3 =  -\beta\gamma-1/2\alpha_+\partial\phi-1/2 cb,  \\
&& j^+ = b-\beta c, \\
&& j^- = \gamma(b-\beta c)-\alpha_+\partial\phi c+(2k+1)\partial c,
\eeqa
and
\beqa
&& {\cal D}^- = -\delx,                       \label{dminus} \\
&& {\cal D}^+ = x^2\delx-2jx+x\theta\delt, \\
&& {\cal D}^3 = -x\delx+j-1/2\theta\delt,     \label{dplus} \\
&& {\cal G}_{-1/2} = \delt - \theta\delx,      \\
&& {\cal G}_{1/2} = x(\delt-\theta\delx)+2j\theta, 
\eeqa
where the conformal weights of fermions $(b,c)$ and bosons 
$(\beta, \gamma)$ are $(1,0)$. The above $osp(1,2)$ currents 
$(J^a, j^b)$ are translated into $({\cal D}^a, {\cal G}_b)$ by the 
replacement $(\beta,\gamma)\ra(\delx,x)$, $(b,c)\ra(\delt,\theta)$ 
and ${1\over2}\alpha_+\partial\phi \ra -j$. In this translation, 
the $c\partial c$ and $\partial\gamma$ terms should be regarded as 
quantum corrections which cancel a redundant pole structure of OPEs.

Since the first type operators resemble eqs.\eq{dminus}-\eq{dplus} 
(for simplicity we discuss the $sl(2)$ part only), 
\beqa
&& L_{-1} = -q^{-1}\delx, \\
&& -L_1 = q^{-1}x^2\delx + q^{-1} x\theta\delt, \\
&& L_0 = -q^{-1}x\delx - {1\over[2]}\theta\delt, 
\eeqa
we assume the following forms from the translation prescription 
\beqa
&& J^+ = -q^{-1}\beta,    \\
&& J^- = q^{-1}\gamma^2\beta + q^{-1}\gamma cb -k_q\partial\gamma 
      +(k+1)_q c\partial c + {\tilde\alpha}_+\gamma\partial\phi,\\
&& J^3 = -q^{-1}\gamma\beta-{1\over[2]}cb -A\alpha_+\partial\phi,
\eeqa
where $k_q$, $(k+1)_q$, $A$, ${\tilde\alpha}_+$, $\alpha_+$ are 
constants, which will be fixed later. The correction terms and a 
scalar field $\phi$ are added. Hereafter all the free fields should 
be regarded as deformed fields. 
The commutation relations of the free fields are translated 
from those of the quantum space \eq{qspace}: For $z\not=w$, 
\beq
c(z)^2 =b(z)^2=0, \quad c(z)\gamma(w)=q^{-1}\gamma(w)c(z),
      \quad \beta(z)b(w)=q^{-1}b(w)\beta(z),  \label{quantumbc}
\eeq
\beqa
&& \beta(z)\gamma(w)=q^{-2}\gamma(w)\beta(z),
                      \qquad b(z)c(w)=q^{-2}c(w)b(z),   \\
&& \beta(z)c(w)=q^{-1}c(w)\beta(z),
   \qquad b(z)\gamma(w)=q^{-1}\gamma(w)b(z), \label{qbetac}
\eeqa
whereas $\beta$ ($b$) and $\gamma$ ($c$) (anti-)commute each other 
at the same point $z=w$,
\beq
           \beta(z)\gamma(z)=\gamma(z)\beta(z), 
          \qquad b(z)c(z)=-c(z)b(z).            \label{unity}
\eeq
Eqs.\eq{unity} correspond to the 1's which appear in the original 
commutation relations \eq{qspace}. Note also that we have slightly 
modified the $c\gamma$ and $bc$ relations (c.f. eq.\eq{qspace}).
Using these relations, we verify the following
\beqa
&& J^+(z)J^-(w) = q^{-4}J^-(w)J^+(z),  \label{j+j-} \\
&& J^3(z)J^-(w) = q^{-2}J^-(w)J^3(z),  \label{j3j-} \\
&& J^3(z)J^+(w) = q^{2}J^+(w)J^3(z),   \label{j3j+} 
\eeqa
which are exact relations up to the correction term $c\partial c$. 
If we want entirely holding relation including the correction term, 
we have to assume $(k+1)_q$ to be a {\it quantum} constant which 
satisfies
\beq
cb(k+1)_q = q^2 (k+1)_q cb,
\eeq
and the correction term $c\partial c$ is then harmless to \eq{j3j-}. 
Although we do not impose such quantum relations on other constants 
anymore, this concept is in common with quantum matrix group theory.
The modification, mentioned below \eq{unity}, of $c\gamma$ and $bc$ 
relations is necessary to eq.\eq{j3j-}. 

Let us investigate the pole structures of the deformed $sl(2)$ 
currents. In the first place, we need to know the pole structures of 
the free fields. From consistency with \eq{qspace}, of course as well 
as with the commutation relations \eq{quantumbc}-\eq{qbetac}, 
the pole structures of the free fields can be determined 
\beqa
&& \beta(z)\gamma(w)={1\over z-w}, 
                     \qquad \gamma(z)\beta(w)={-q^2\over z-w}, \\
&& b(z)c(w)={1\over z-w}, \qquad c(z)b(w)={q^2\over z-w},
\eeqa
and we assume
\beq
\phi(z)\phi(w)=+log(z-w). 
\eeq
We adopt the commutation relations of $\phi$ 
such that the relations \eq{j+j-} and \eq{j3j-} hold
\beq
\phi(z)\beta(w)=q^2\beta(w)\phi(z),
               \qquad\phi(z)\gamma(w)=q^{-2}\gamma(w)\phi(z).
\eeq

Now, we can calculate OPEs among the currents and verify the 
relations \eq{j+j-}-\eq{j3j+} at the level of pole structures (up to 
regular terms). We must note the following rule for power contractions
when performing Wick contractions:
\beq
\beta(z)\gamma(w)^n =
(1+q^{-2}+.....+q^{-2(n-1)}){1\over z-w}\gamma(w)^{n-1}
={q^{-n+1}[n]\over z-w}\gamma(w)^{n-1},
\eeq 
and the similar rule applies to a product of $b$ $c$ fields, 
for example,
\beq
b(z)c(w)\partial c(w) = {1\over z-w}\partial c(w) 
                       -{q^{-2}\over (z-w)^2}c(w).
\eeq
Keeping these rules in mind, we finally obtain
\beqa
&& q^{-1}J^3(z)J^+(w)=qJ^+(w)J^3(z)={1\over z-w}J^+(w), \label{jj1}\\
&& qJ^3(z)J^-(w)=q^{-1}J^-(w)J^3(z)={-1\over z-w}J^-(w), \\
&& q^{2}J^+(z)J^-(w)=q^{-2}J^-(w)J^+(z)={[2]\over z-w}J^3(w), 
\label{jj3}\\
&& J^3(z)J^3(w)={[2]^{-1}d \over (z-w)^2, }         \label{jj4}   
\eeqa
where $A$, $d$ and $k_q$ are fixed  
\beqa
&& A = {q\over [2]}, \\
&& d = -[2]+{q^2\over[2]}+{q^2\over[2]}\alpha_+^2,  \\ 
&& k_q = -[2]+{1\over[2]}+{1\over[2]}{\tilde\alpha}_+\alpha_+=0. 
\eeqa
Here, $k_q=0$ is imposed in order that the first equality of 
\eq{jj3} holds concerning higher singularity. It can be shown from 
\eq{jj1}-\eq{jj4} that the mode expansions of the deformed currents
\beq
J^a(z) = \sum_{n\in Z}J^a_n z^{-n-1}
\eeq
satisfy the CZ type deformation of affine $sl(2)$ algebra
\beqa
&& [J^+_n,J^-_m]_{(2,-2)}= [2]J^3_{n+m}, \\
&& [J^3_n,J^+_m]_{(-1,1)}=   J^+_{n+m}, \\
&& [J^3_n,J^-_m]_{(1,-1)}= - J^-_{n+m}, \\
&& [J^3_n,J^3_m]={n\over[2]}d\delta_{n+m,0}.
\eeqa                                                                  
The situation of $d$ and $k_q$ considerably differs from conformal 
field theory ($q=1$ case). In the case of $q=1$, we have $d=k_q$, 
and $k_q$ is not zero generally. Let us put $d=k_q$. Irrespectively 
of the value of $k_q$, we find 
\beq
     {\tilde\alpha}_+ = q^2\alpha_+ + (1-q^2)\alpha_- ,
\eeq
where $\alpha_+\alpha_-=-1$ is used \cite{DF}. It may be interesting 
that the coefficient of the correction term $\gamma\partial\phi$ is 
given by an average of screening charges $\alpha_{\pm}$. 

Finally, let us observe the case of non-super quantum space. 
Dropping $b$ and $c$ fields from the above arguments, we get 
\beq
d=-[2]+{q^2\over[2]}\alpha_+^2, \qquad 
k_q=-[2]+{1\over[2]}{\tilde\alpha}_+\alpha_+=0, 
\eeq
and $d=k_q$ means 
\beq
     {\tilde\alpha}_+ = q^2\alpha_+ .
\eeq
In this case, the quantum differential operators corresponding to 
${\cal D}^a$ ($a=\pm,3$) are 
\beqa
&& D^- = -q^{-1}\delx, \\
&& D^+ = q^{-1}x^2\delx -q^{-1}x(2j)_q,   \label{quantumD}\\
&& D^3 = -q^{-1}x\delx + {1\over[2]}(2j)_q, 
\eeqa
where $(2j)_q$ is again the quantum constant such that
\beq
x(2j)_q = q^2 (2j)_q x, \qquad  \delx(2j)_q = q^{-2} (2j)_q \delx.
\eeq
$D^a$ satisfy the following commutation relations, which are 
essentially the same as the subalgebra of CZ algebra.
\beqa
&& [D^+, D^-]_{(-2,2)} = -[2]D^3, \\
&& [D^3, D^+]_{(1,-1)} = - D^+, \\
&& [D^3, D^-]_{(-1,1)} =   D^-.
\eeqa

\section{Conclusion}
\setcounter{equation}{0}
\indent                     

We have presented the general construction method of $N=1$ super CZ 
algebra based on quantum superspace differential operators, after 
analyzing two types of superalgebra. Then we have shown a few 
examples of its subalgebra realizations in the $q$-oscillator and in 
conformal field theoretical analogy, where quantum space structure 
was crucial.  

Although we did not specify abstract and unique commutation 
relations on super generator parts, we have found several useful and 
universal structures of super CZ algebras as discussed in sect. 4. 
First of all, the CZ generators can always be given in automatic way 
by the base elements ${\cal B}_n$ and ${\cal F}_n$, which are nothing 
but bilinear combinations of $G_r$ operators. This means that we do 
not have to take care of the CZ generators at the beginning stage of 
supersymmetric extension. All what we have in mind is the only algebra 
of $G_r$. Namely, the $G_r$ algebra determines the whole structure of 
super CZ algebras. This situation is similar to the undeformed case. 
The remaining key point is how to define convenient $G_r$ operators 
which will lead to simple commutation relations.

Secondly, our super CZ algebras acquire non-linear 
deformation terms since the ${\cal F}_n$ corresponds to a non-linear 
deformation term in the superalgebra. On the other hand, 
${\cal B}_n$ corresponds to a linear term of the superalgebra. The 
introduction of non-linear deformation (base elements $B_n$ and 
$F_n$) is non-trivial because of no separation between $B_n$ and 
$F_n$ in the case of $q=1$. As a result, this base structure 
enables the super CZ algebras to be closed forms. 

Following to these structures, we have become able to systematically 
approach a supersymmetric extension of the CZ algebra. Clearly, 
our method is better than a blind calculation. This is a big step 
to examine many other possibilities which arise from various 
definitions of $G_r$ operators and from other quantum superspace 
commutators \cite{KU}. 

The non-linear terms are originated in the complexity of {\it super} 
space algebra structure, and hence we have the non-trivial question 
whether or not a linearized simple super CZ algebra is possible 
based on the quantum superspace construction. Unfortunately, we have 
not got yet any answer to this question. We do not know yet which 
superalgebra is simple and suitable for arguing abstract algebraic 
features like the CZ generator constraints. If this could be solved, 
the similar structure as the embedding of CZ generators into the 
Lie algebra type deformation \cite{embed} might be found for the 
superalgebra. Furthermore, the super Lie algebra type deformations 
of Virasoro algebra \cite{superV} might be realized in quantum 
superspace operators (some of them were already realized in 
non-quantum space $q$-differential operators \cite{saito}).

We restricted our interests to $N=1$ deformations in this report. 
However, all/some of these problems appeared in this paper could be 
solved if we consider a $N=2$ extension or two-parameter deformation 
of quantum superspace. Both possibilities are natural options 
because $F_n$ is $N=2$ algebra element, and there is no reason of 
introducing a common deformation parameter with bosonic and 
fermionic spaces.

In connection with some physical arguments, it would be interesting 
to discuss about the possibility that the non-commutative nature of 
quantum space coordinates could be related to a discrete physical 
system in a similar sense to matrix-model-like argument. Also, we 
should consider what else would be behind the idea of quantum groups 
and quantum (super)spaces. For example, $q=0$ quantum group is an 
interesting subject \cite{aref} in connection with the master field 
algebra of large $N$ matrix model. Of course, those who wish 
to apply deformed algebras to integrable systems should find Hopf 
algebra structures. 
                                                                      
Although we analyzed only a few examples, we believe that our 
method and formulas will be a useful foundation toward physical 
applications of quantum superspaces and the establishment of 
abstract algebraic results for super CZ algebras.

%


\begin{thebibliography}{99}
%
\bibitem{QG} "Yang-Baxter Equation in Integrable Systems", ed. M. Jimbo 
(World Scientific, Singapore, 1990).
\bibitem{qos} A.J. Macfarlane \JP{A22} (1989) 4581;\\
              L.C. Biederharn, \JP{A22} (1989) L873;\\
              T. Hayashi, \CMP{127} (1990) 129.
\bibitem{Pauli} R.N. Mohapatra, \PL{B242} (1990) 407.
\bibitem{regge} M. Chaichian, J.F. Gomes and R. Gonzalez Felipe, 
                \PL{B341} (1994) 147.
\bibitem{optic} M. Chaichian, D. Ellinas and P. Kulish, \PRL{65} (1990) 980.
\bibitem{Fuji} K. Fujikawa, L.C. Kwek and C.H. Oh, \MPL{A10} (1995) 2543.
%
\bibitem{CZ} T. Curtright and C. Zachos, \PL{B243} (1990) 237.
\bibitem{zha} C.-Z. Zha, \JMP{35} (1994) 517.
\bibitem{jev} A. Jevicki and A. van Tonder, \MPL{A11} (1996) 1397.
\bibitem{fermi} G.W. Semenoff and R.J. Szabo, hep-th/9605140.
\bibitem{QSST} T. Kobayashi and T. Uematsu, \PL{B306} (1993)27; 
               \ZPC{56} (1992) 193. 
\bibitem{WZ} J. Wess and B. Zumino, {\it Nucl. Phys.} (Proc. Suppl.) 
            {\bf 18B} (1990) 302. 
\bibitem{arch}I.V. Volovich, "Interacting p-adic fields and 
non-commutative geometry", lecture XIXth Conference on Differential 
geomotric methods in theoretical physics (Rappalo, 1990).
\bibitem{qregular} S. Majid, \IJMP{A5} (1990) 4689.
\bibitem{qgauge} T. Brzezinski and S. Majid, \CMP{157} (1993) 591.
\bibitem{qscalar} M.R. Ubriaco, \MPL{A9} (1994) 1121.
\bibitem{cova} M. Chaichian, P. Kulish and J. Lukierski, \PL{B262} (1991) 43.
\bibitem{KU} T. Kobayashi, \ZPC{60} (1993) 101. 
\bibitem{FMS} D. Friedan, E. Martinec and S. Shenker, \NP{B271} (1986) 93.
\bibitem{hass1}A.El Hassouni, Y. Hassouni, E.H. Tahri and M. Zakkari,
                 \MPL{A10} (1995) 2169.
\bibitem{hass2}  {\it idem.} \MPL{A11} (1996) 37. 
\bibitem{AS} H. Sato, \PTP{89} (1993) 531; \\
         N. Aizawa and H. Sato, \PL{B256} (1991) 185. 
\bibitem{osp12} P. Kulish and N.Yu. Reshetikhin,\LMP{18} (1989) 143;\\ 
         M. Chaichian, P. Kulish and J. Lukierski, \PL{B237} (1990) 401.
\bibitem{osp} H. Saleur, \NP{B336} (1990) 363. 
\bibitem{map} R. Floreanini and L. Vinet, \JP{A23} (1990) L1019.
\bibitem{BO} M. Bershadsky and H. Ooguri, \PL{B229} (1989) 374.
\bibitem{DF} VI.S. Dotsenko and V.A. Fateev, \NP{B240} (1984) 312;
             {\it ibid.} {\bf B251} (1985) 691.
\bibitem{embed} H. Sato, \NP{B393} (1993) 442; \\
              M. Chaichian and P.P. Presnajder, hep-th/9603064.
\bibitem{superV} M. Chaichian and P.P. Presnajder, \PL{B277} (1992) 109;\\
          A.A. Belov and K.D. Chaltikian, \MPL{A8} (1993) 1233;\\
          H-T. Sato, \NP{B471} (1996) 553;\\
          E. Batista, J.F. Gomes and I.J. Lautenschleguer, q-alg/9603004. 
\bibitem{saito} R. Kemmoku and S. Saito, \PL{B319} (1993) 471;
hep-th/9411027.
\bibitem{aref} L. Accardi, I.Ya. Aref'eva, S.V. Kozyrev and I.V. Volovich, 
\MPL{A10} (1995) 2323;\\
 I.Ya. Aref'eva, and I.V. Volovich, \NP{B462} (1996) 600.
\end{thebibliography}
\end{document}